\documentclass[3p,times,twocolumn]{elsarticle}

\usepackage{ecrc}

\volume{00}

\firstpage{1}

\journalname{Nuclear and Particle Physics Proceedings}

\runauth{M. Koles\'{a}r, J. Novotn\'y}

\jid{nppp}

\jnltitlelogo{Nuclear and Particle Physics Proceedings}

\usepackage{amssymb}
\usepackage[figuresright]{rotating}

\newcommand{\no}{\noindent}
\newcommand{\vsp}[1]{\vspace{#1}}
\newcommand{\hsp}[1]{\hspace{#1}}
\newcommand{\Op}{\mathcal{O}}
\newcommand{\myeq}[3]{\vspace{#2} \begin{equation} \hspace{#1} #3 \end{equation} \vspace{0cm}}

\begin{document}

\begin{frontmatter}



\dochead{}

\title{Convergence properties of $\eta$$\,\to$$3\pi$ in low energy QCD}


\author[label1]{Mari\'{a}n Koles\'{a}r\corref{cor1}}
\address[label1]{Institute of Particle and Nuclear Physics, Faculty of Mathematics and Physics, Charles University, Prague, Czech republic}
\cortext[cor1]{Speaker}
\author[label1]{Ji\v{r}\'i Novotn\'y}

\begin{abstract}
Even today, the convergence of the decay widths and some of the Dalitz plot parameters of the $\eta $$\,\rightarrow $$3\pi$ decays seems problematic in low energy QCD. We provide an overview of the current experimental and theoretical situation with historical background and summarize our recent results, which explore the question of compatibility of experimental data with a reasonable convergence of a carefully defined chiral series in the framework of resummed chiral perturbation theory.
\end{abstract}

%% \begin{keyword}


%% \end{keyword}

\end{frontmatter}

% \linenumbers


\section{Overview}

From the very beginning it was understood that the $\eta $$\,\rightarrow $$3\pi 
$ decays are isospin breaking processes. Initially, the decays were thought to be of electromagnetic origin \cite%
{Bose:1966,Bardeen:1967}, generated by the isospin breaking virtual photon
exchange

\begin{equation}
H_{QED}(x)=-\frac{e^{2}}{2}\int dyD^{\mu \nu }(x-y)T(j_{\mu }(x)j_{\nu }(y)).
\label{H_QED}
\end{equation}

\no 
Simultaneously, however, it was discovered that the decays are almost
forbidden in the framework of QED (the Sutherland theorem \cite
{Sutherland:1966zz,Bell:1968mi}), which was met with some disbelief \cite{Bardeen:1967}. 

The early calculations \cite{Bose:1966,Bardeen:1967}, applying current algebra and PCAC, related the $\eta $-$\pi $ matrix elements to the difference of squared kaon masses or kaon and pion masses, respectively. In
fact, the latter resembles the later Dashen's theorem, which cannot be justified by
electrodynamics \cite{Bell:1968mi}. In spite of that, the obtained value for the neutral channel decay rate were of approximately correct order of magnitude, $\Gamma^0$=160 eV.

Hence it became apparent that there
has to be a source of isospin breaking beyond the term (\ref{H_QED}) \cite%
{Osborn:1970nn}. As is known today, strong interactions break isospin via the
difference between the masses of the $u$ and $d$ quarks

\begin{equation}
H_{QCD}^{IB}(x)=\frac{m_{d}-m_{u}}{2}(\bar{d}(x)d(x)-\bar{u}(x)u(x)).
\end{equation}

The work \cite{Osborn:1970nn} collected all the relevant current
algebra terms contributing to the decays and can be considered to be
the first to provide the correct leading order calculation. The
obtained value for the neutral decay rate did not significantly change though and turned out to be much lower than the
experimental value then available ($\Gamma^0$=164 eV vs 750$\pm$200 eV). There remained a significant discrepancy, as was concluded in \cite{Osborn:1970nn}.

When a systematic approach to low energy hadron physics was born in the form
of chiral perturbation theory ($\chi $PT) \cite%
{Weinberg:1978kz,Gasser:1983yg,Gasser:1984gg}, it was quickly applied to the 
$\eta$$\,\rightarrow$$3\pi $ decays \cite{Gasser:1984pr}. The one loop
corrections were very sizable, the result for the decay width of the charged
channel was $\Gamma^+$=160$\pm $50 eV, compared to the current algebra prediction of 66
eV. However, already at that time there were hints that the experimental
value is still much larger (340$\pm $100 eV), thus ``resurrecting the puzzle" the theory aimed to solve. 
The current PDG value \cite{PDG:2014kda} is 

\myeq{1.75cm}{0cm}{\Gamma^+_\mathrm{exp} = 300 \pm 12 \ \mathrm{eV}.}

\no In the case of the neutral channel, the average is \cite{PDG:2014kda}

\myeq{1.75cm}{0cm}{\Gamma^0_\mathrm{exp} = 428 \pm 17 \ \mathrm{eV}.}

After the effective theory was extended to include virtual photon exchange
generated by (\ref{H_QED}) \cite{Urech:1994hd}, it was shown that the
next-to-leading electromagnetic corrections to the Sutherland's theorem are
very small as well \cite{Baur:1995gc, Ditsche:2008cq}. The theory thus seems to converge really slowly for the decays. At last, the
two loop $\chi $PT calculation \cite{Bijnens:2007pr} has succeeded to
provide a reasonable prediction for the decay widths. 

Meanwhile, experimental data are being gathered with increasing precision in order to
make more detailed analysis of the Dalitz plot distribution possible.
Comparison of the recent experimental information with the NNLO $\chi $PT
results can be found in tables \ref{tab1} and \ref{tab2}, with the conventionally
defined Dalitz plot parameters defined as

\myeq{-0.5cm}{0cm}{
	\eta\to\pi^0\pi^+\pi^-:\,\ |A|^2 = A_0^2 (1+ay+by^2+dx^2+\dots)\quad }
\myeq{-0.5cm}{-0.25cm}{
	\eta\to 3\pi^0:\qquad |\overline{A}|^2 = \overline{A}_0^2(1+\alpha z+\dots),}

\no where $x$$\,\sim$$\,u$$-$$t$, $y$$\,\sim$$\,s_0$$-$$s$, $z$$\,\sim$$\,x^2$+$y^2$ and 
$s_0$ is the Dalitz plot center $s_0=1/3(M_\eta^2$+$3M_\pi^2)$.
For the sake of brevity, we added the systematic and statistical uncertainties in squares.
As can be seen, a tension between $\chi $PT and experiments appears to
be in the charged decay parameter $b$ and the neutral decay parameter $\alpha$.

Alternative approaches were developed in order to model the amplitudes more
precisely, namely dispersive approaches \cite%
{Kambor:1995yc,Anisovich:1996tx,Colangelo:2011zz,Kampf:2011wr,Guo:2015zqa} and
non-relativistic effective field theory \cite%
{Bissegger:2007yq,Gullstrom:2008sy,Schneider:2010hs}. These more or less
abandon strict equivalence to $\chi $PT and succeed in reproducing a
negative sign for $\alpha $ (see table \ref{tab2}) 

Thence comes our motivation to ask whether it is possible to carefully define an amplitude with reasonable convergence properties which would reproduce the experimental data for the decay widths and the Dalitz plot parameters. In other words, we aim to investigate the question of compatibility of the experimental data with a reasonable convergence of the chiral series.

\hsp{-0.5cm}
\begin{table}[t]
\begin{center}
{\footnotesize
\begin{tabular}{|c|c|c|c|}
\hline
\rule[-0.2cm]{0cm}{0.6cm} $\eta\to\pi^+\pi^-\pi^0$ & $a$ & $b$ & $d$ \\ 
\hline
\rule[-0.2cm]{0cm}{0.6cm} Cr.Barrel '98 \cite{CrBarrel:1998yi} & 
$-1.22\pm0.07$ & $0.22\pm0.11$ & $0.06$ (input)\\ 
\rule[-0.2cm]{0cm}{0.5cm} KLOE '08 \cite{KLOE:2008ht} & $-1.090\pm0.020$ & $%
0.124\pm0.012$ & $0.057\pm0.017$\\
\rule[-0.2cm]{0cm}{0.5cm} WASA '14 \cite{Wasa:2014aks} & $-1.144\pm0.018$ & $%
0.219\pm0.042$ & $0.086\pm0.025$ \\
\rule[-0.2cm]{0cm}{0.5cm} BESIII '15 \cite{BESIII:2015cmz} & $-1.128\pm0.017$ & $%
0.153\pm0.017$ & $0.085\pm0.018$ \\
\rule[-0.2cm]{0cm}{0.5cm} KLOE '16 \cite{KLOE:2016qvh} & $-1.095\pm0.004$ & $%
0.145\pm0.006$ & $0.081\pm0.007$ \\ \hline
\rule[-0.2cm]{0cm}{0.6cm} NREFT '11 \cite{Schneider:2010hs} & $%
-1.213\pm0.014 $ & $0.308\pm0.023$ & $0.050\pm0.003$ \\ 
\rule[-0.2cm]{0cm}{0.5cm} NNLO $\chi$PT '07 \cite{Bijnens:2007pr} & $%
-1.271\pm0.075$ & $0.394\pm0.102$ & $0.055\pm0.057$ \\ \hline

\end{tabular}
}
\end{center}
\caption{Recent experimental and theoretical results for $\eta\to\pi^+\pi^-\pi^0$.}
\label{tab1}
\end{table}

\begin{table}[t]
\begin{center}
{\normalsize {\footnotesize
\begin{tabular}{|c|c|}
\hline
\rule[-0.2cm]{0cm}{0.6cm} $\eta\to\pi^0\pi^0\pi^0$ & $\alpha$ \\ \hline
\rule[-0.2cm]{0cm}{0.6cm} Crystal Barrel '98 \cite{CrBarrel:1998yj} & $%
-0.052\pm0.020$ \\ 
\rule[-0.2cm]{0cm}{0.5cm} SND '01 \cite{Achasov:2001xi} & $-0.010\pm0.023$
\\ 
\rule[-0.2cm]{0cm}{0.5cm} Crystal Ball '01 \cite{CrBall:2001fm} & $%
-0.031\pm0.004$ \\ 
\rule[-0.2cm]{0cm}{0.5cm} CELSIUS/WASA '07 \cite{Wasa:2007aa} & $%
-0.026\pm0.014$ \\ 
\rule[-0.2cm]{0cm}{0.5cm} WASA at COSY '09 \cite{Wasa:2008vn} & $%
-0.027\pm0.009$ \\ 
\rule[-0.2cm]{0cm}{0.5cm} Crystal Ball at MAMI-B '09 \cite{CrBall:2008ny} & $%
-0.032\pm0.003$ \\ 
\rule[-0.2cm]{0cm}{0.5cm} Crystal Ball at MAMI-C '09 \cite{CrBall:2008ff} & $%
-0.0322\pm0.0025$ \\ 
\rule[-0.2cm]{0cm}{0.5cm} KLOE '10 \cite{KLOE:2010mj} & $-0.0301\pm0.0050$
\\ 
\rule[-0.2cm]{0cm}{0.5cm} PDG '14 \cite{PDG:2014kda} & $-0.0315\pm0.0015$ \\ 
\hline
\rule[-0.2cm]{0cm}{0.6cm} NREFT '11 \cite{Schneider:2010hs} & $%
-0.0246\pm0.0049 $ \\ 
\rule[-0.2cm]{0cm}{0.5cm} Prague disp.fit '11 \cite{Kampf:2011wr} & $%
-0.044\pm0.004$ \\ 
\rule[-0.2cm]{0cm}{0.5cm} Bern disp.fit '11 \cite{Colangelo:2011zz} & $%
-0.045\pm0.010$ \\
\rule[-0.2cm]{0cm}{0.5cm} Guo et al. '15 \cite{Guo:2015zqa} & $%
-0.022\pm0.004$ \\
\rule[-0.2cm]{0cm}{0.5cm} NNLO $\chi$PT '07 \cite{Bijnens:2007pr} & $%
+0.013\pm0.032$ \\ \hline
\end{tabular}
} }
\end{center}
\caption{Recent experimental and theoretical results for $\eta\to 3\pi^0$. }
\label{tab2}
\end{table}

\vsp{-0.5cm}
\section{Calculation}

There is a long standing suspicion that chiral perturbation theory might
posses slow or irregular convergence in the case of the three light quark
flavours \cite{Fuchs:1991cq,DescotesGenon:1999uh}. An alternative method,
now dubbed resummed $\chi $PT \cite%
{DescotesGenon:2003cg,DescotesGenon:2007ta}, was developed in order to incorporate such a possibility. The
starting point is the realization that the standard approach to $\chi $PT,
as a usual treatment of perturbation series, implicitly assumes good
convergence properties and hides the uncertainties associated with a
possible violation of this assumption. The resummed procedure uses the same
standard $\chi $PT Lagrangian and power counting, but only expansions
derived linearly from the generating functional are considered safe. All subsequent
manipulations are carried out in a non-perturbative algebraic way. The
expansion is done explicitly to next-to-leading order and higher orders are
collected in remainders. These are not neglected, but retained as sources of
error, which have to be estimated.

The working
hypothesis of the resummed approach is that only a limited set of safe observables, as defined above, has the property of global convergence, i.e. that the NNLO remainders are of a natural order of magnitude. Observables derived from the safe ones by means of nonlinear relations do not in general satisfy the criteria for global convergence due
to the possible irregularities of the chiral series. Therefore, it is
necessary to express such dangerous observables in terms of the safe ones in a non-perturbative way.

Our calculation is described in depth in \cite{Kolesar:2016jwe}. What we present here is
only a brief excerpt, meant as a summary of the basic steps and obtained results. 

Within the formalism, we start by expressing the charged decay amplitude in terms of the 4-point
Green functions $G_{ijkl}$. We compute at first order in isospin braking. 
In this case the amplitude takes the form

\myeq{-0.5cm}{0cm}{
	F_\pi^3F_{\eta}A(s,t,u)
		= G_{+-83}-\varepsilon_{\pi}G_{+-33}+\varepsilon_{\eta}G_{+-88} + \Delta^{(6)}_{G_D},\ }
		
\no where $\Delta^{(6)}_{G_D}$ is the direct higher order remainder to the complete 4-point Green function. 
The physical mixing angles to all chiral orders and first in isospin braking
can be expressed in terms of quadratic mixing terms of the generating functional to NLO
and related indirect remainders

\myeq{-0.5cm}{0cm}{
	\varepsilon_{\pi,\eta} = -\frac{F_{0}^{2}}{F_{\pi^0,\eta}^{2}}
		\frac{(\mathcal{M}_{38}^{(4)}+\Delta_{M_{38}}^{(6)}) - 									
		M_{\eta,\pi^0}^{2}(Z_{38}^{(4)}+\Delta _{Z_{38}}^{(6)})}
		{M_\eta^2-M_{\pi^0}^2}.}

\no In this approximation the neutral decay channel amplitude can be related to the charged
one as

\myeq{0cm}{0cm}{\overline{A}(s,t,u)=A(s,t,u)+A(t,u,s)+A(u,s,t).}

As dictated by the method, $\Op(p^2)$ parameters appear inside loops, while
physical quantities in outer legs. Due to the leading order masses in loops, 
such a strictly derived amplitude has an
incorrect analytical structure, cuts and poles are placed in unphysical positions. To account
for this, we carefully modify the amplitude using a NLO dispersive representation.
The procedure is described in detail in \cite{Kolesar:2016jwe}.   

The next step is the treatment of the low energy constants (LECs). The leading order
ones, as well as the quark masses, are expressed in terms of convenient parameters

\myeq{-0.6cm}{0cm}{
	Z = \frac{F_0^2}{F_{\pi}^2},\ \
	X = \frac{2F_0^2B_0\hat{m}}{F_{\pi}^2M_{\pi}^2},\ \
	r = \frac{m_s}{\hat{m}},\ \
	R = \frac{(m_s-\hat{m})}{(m_d-m_u)},\ \ }
	
\no where $\hat{m}$=$(m_u+m_d)/2$. The standard approach tacitly assumes values of
$X$ and $Z$ close to one, which means that the leading order terms
should dominate the expansion. However, recent fits \cite{Bijnens:2014lea}
indicate much lower values. A possibility of
a non-standard scenario of spontaneous chiral symmetry breaking is thus still open.

At next-to-leading order, the LECs $L_4$-$L_8$ are algebraically reparametrized in terms
of pseudoscalar masses, decay constants and the free parameters $X$, $Z$ and $r$ using chiral expansions of 
two point Green functions, similarly to \cite{DescotesGenon:2003cg}. Because expansions are formally not truncated, 
each generates an unknown higher order remainder.

We still don't have a similar procedure for $L_1$-$L_3$. Therefore we collect
a set of standard $\chi$PT fits \cite{Bijnens:1994ie,Amoros:2001cp,Bijnens:2011tb, Bijnens:2014lea} and by taking their mean and spread, while 
ignoring the much smaller reported error bars, we obtain an estimate of their influence. As
is shown in \cite{Kolesar:2016jwe}, the results depend on these constants only very weakly. The error bands given
below include the estimated uncertainties in $L_1$-$L_3$.

The $O(p^6)$ and higher order LECs, notorious for their abundance, are collected in a relatively smaller number of higher order remainders. We have a direct remainder to the 4-point Green function and eight indirect ones - three
related to each the pseudoscalar masses and the decay constants, two to the mixing angles. 
The last step leading to numerical results is their estimate.
We use an approach based on general arguments about the convergence 
of the chiral series \cite{DescotesGenon:2003cg}, which leads to

\myeq{0.5cm}{0cm}{\Delta_G^{(4)} = (0 \pm 0.3)\,G,\quad \Delta_G^{(6)} = (0 \pm 0.1)\,G,}

\no where $G$ stands for any of our 2-point or 4-point Green functions,
which generate the remainders. This is in principle an assumption. Hence we test the
compatibility of this assumption of a reasonably good chiral convergence of trusted 
quantities with experimental data in a statistical sense.

\section{Summary of results}

Our results depend, besides the remainders, on several free parameters - the chiral condensate, the chiral decay constant, the strange quark mass and the difference of the light quark masses. They are expressed in terms of the parameters $X$, $Z$, $r$ and $R$, respectively. The quark mass parameters have been fixed from lattice QCD averages \cite{Aoki:2013ldr}: $r$=27.5$\pm$0.4 and $R$=35.8$\pm$2.6.

We have treated the uncertainties in the higher order remainders and other parameters statistically and numerically generated a large range of theoretical predictions, which can be confronted with experimental information. Let us stress that at this point our goal is not to provide sharp predictions, as the theoretical uncertainties are large. Nevertheless, in this form, the approach is suitable for addressing questions which might be difficult to ask within the standard framework.

\begin{figure}[p]
\begin{center}
\includegraphics[scale=0.6]{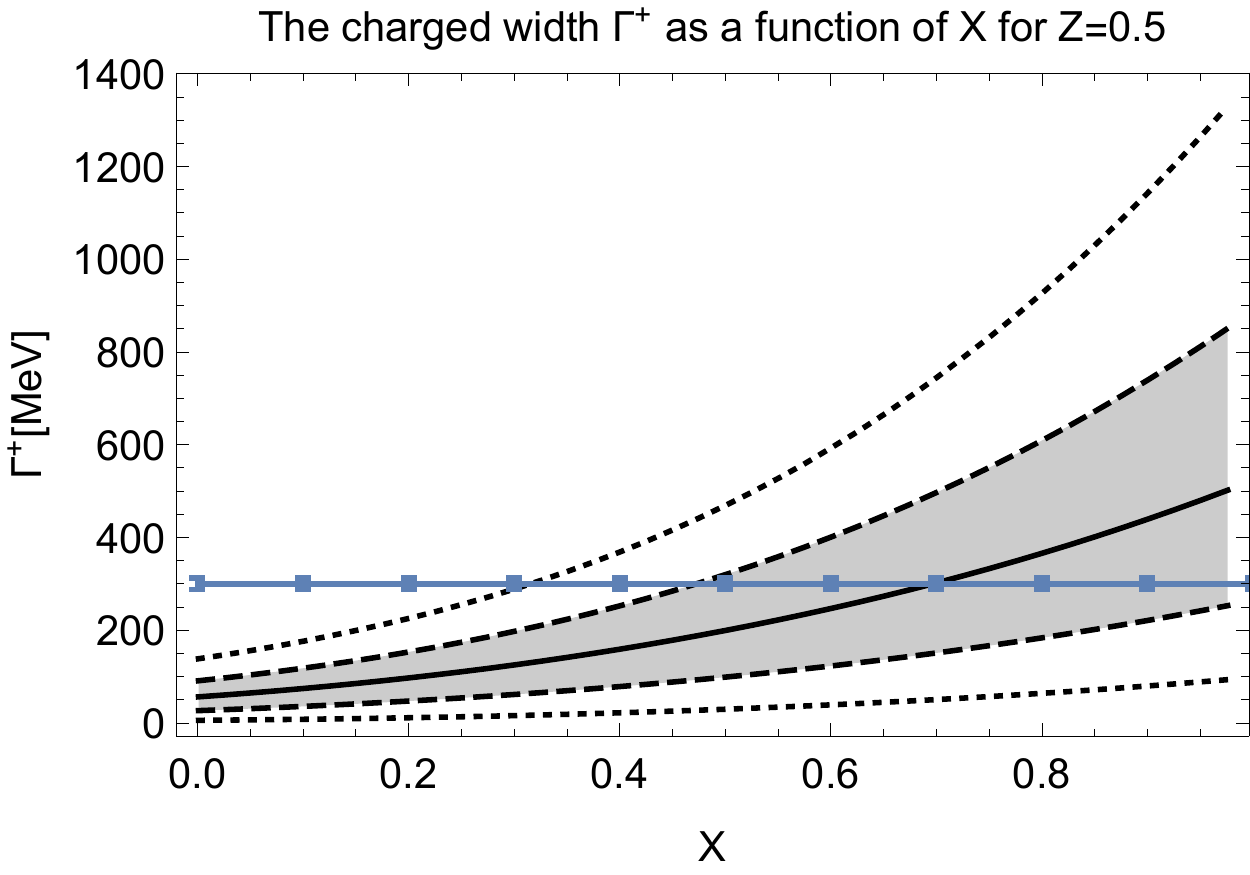}
\includegraphics[scale=0.6]{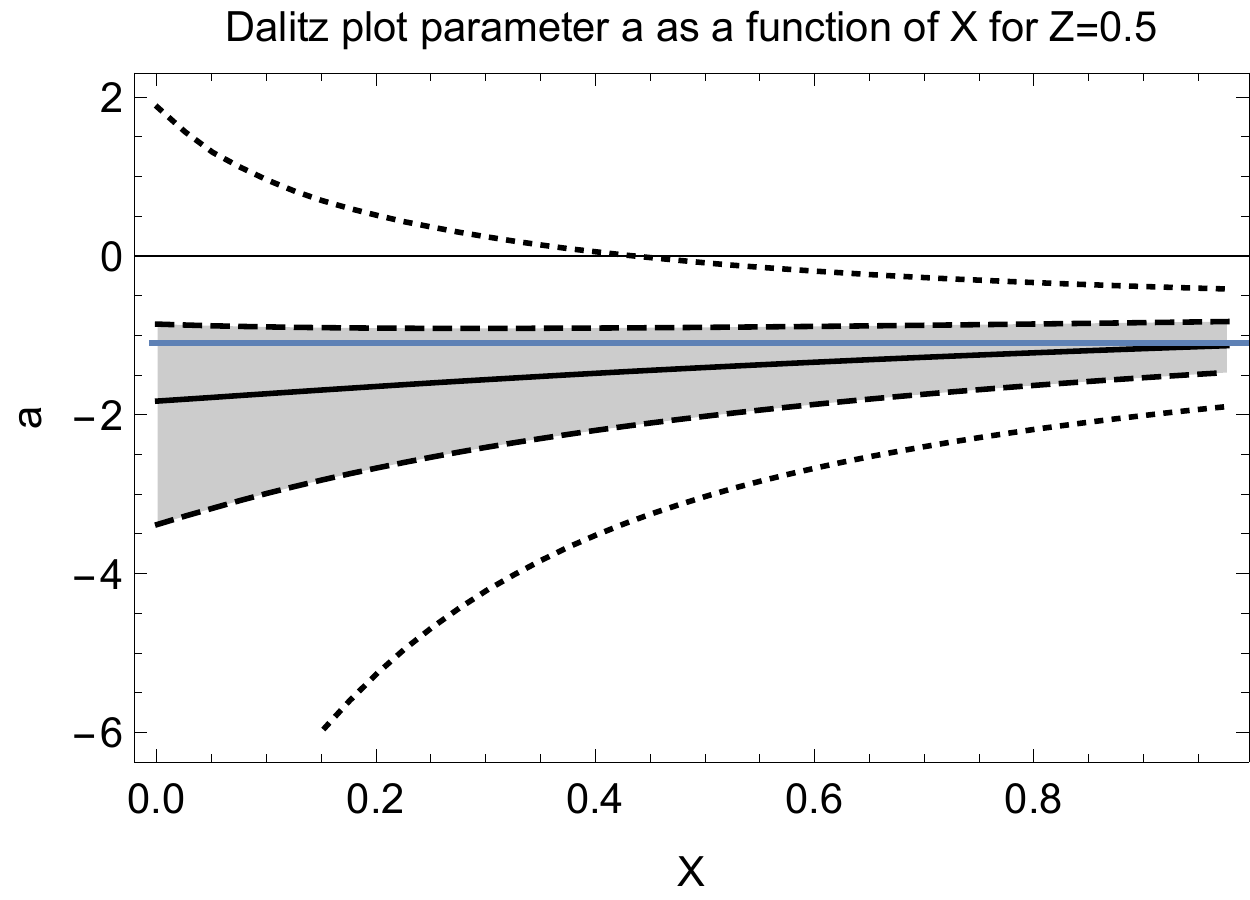}
\includegraphics[scale=0.6]{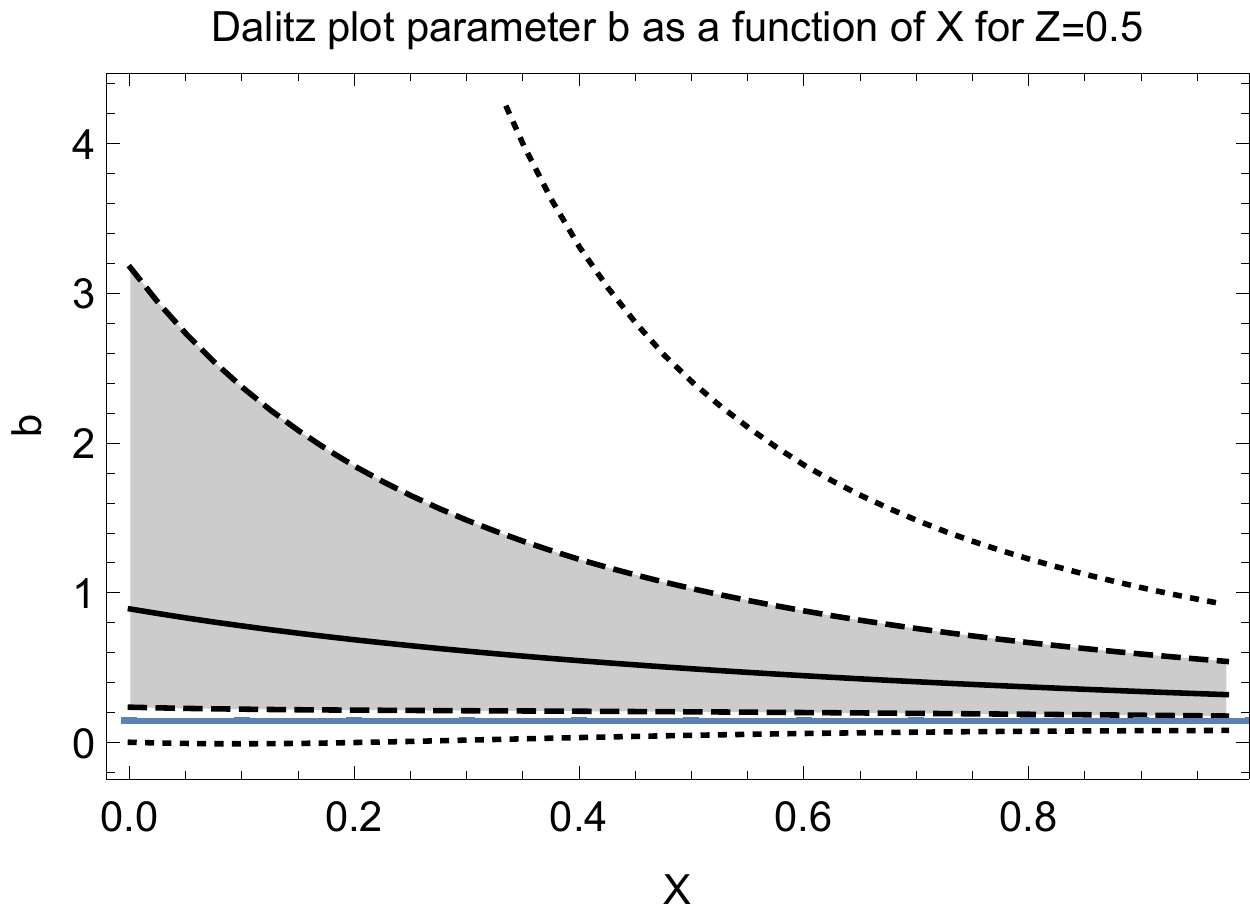} 
\includegraphics[scale=0.6]{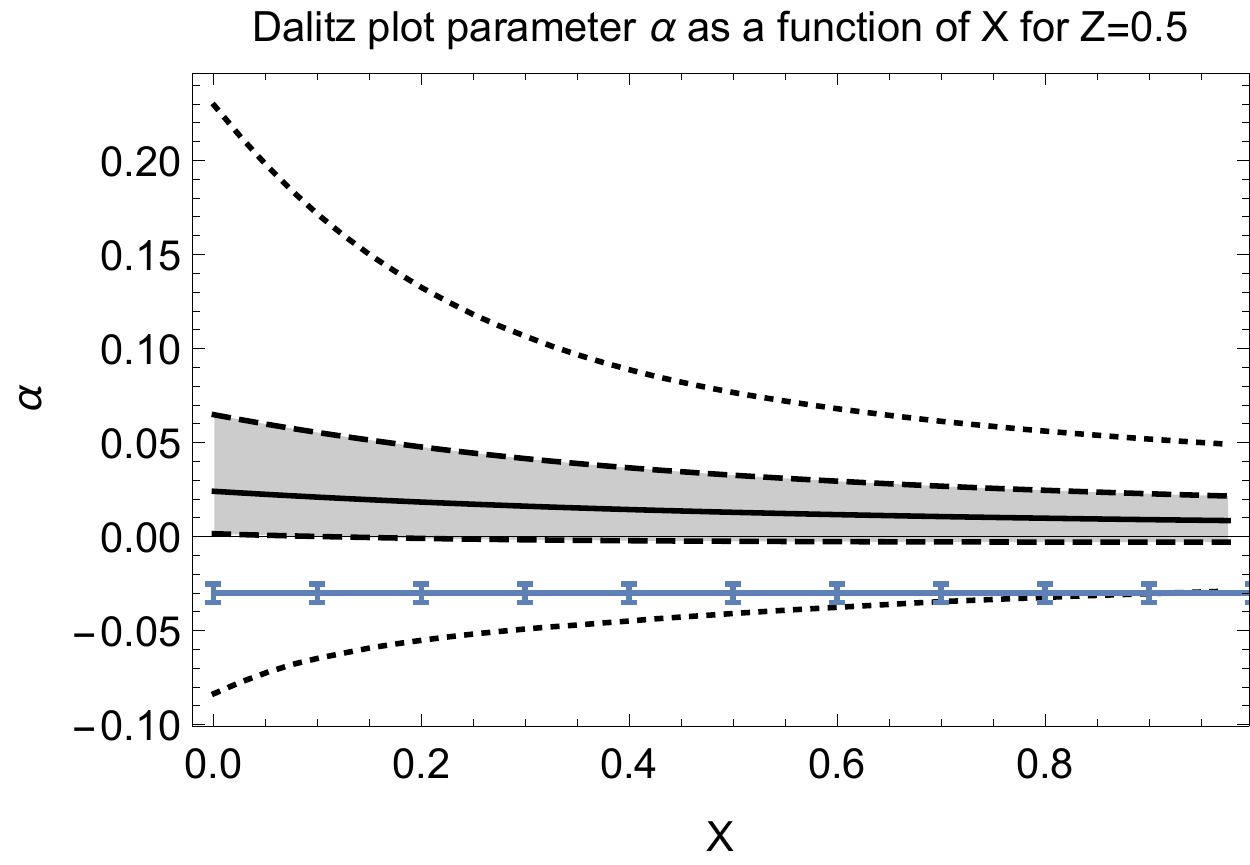}
\end{center}
\caption{Parameters $\Gamma^+$, $a$, $b$ and $\alpha$ as a function of $X$ for $Z=0.5$.\newline
The median (solid line), the one-sigma band (dashed,
shadowed) and the two-sigma band (dotted) are depicted along with the
experimental value \cite{PDG:2014kda,KLOE:2016qvh,KLOE:2010mj} (solid horizontal line with error bars).}
\label{f1}
\end{figure}

\begin{figure}[p]
\begin{center}
\includegraphics[scale=0.6]{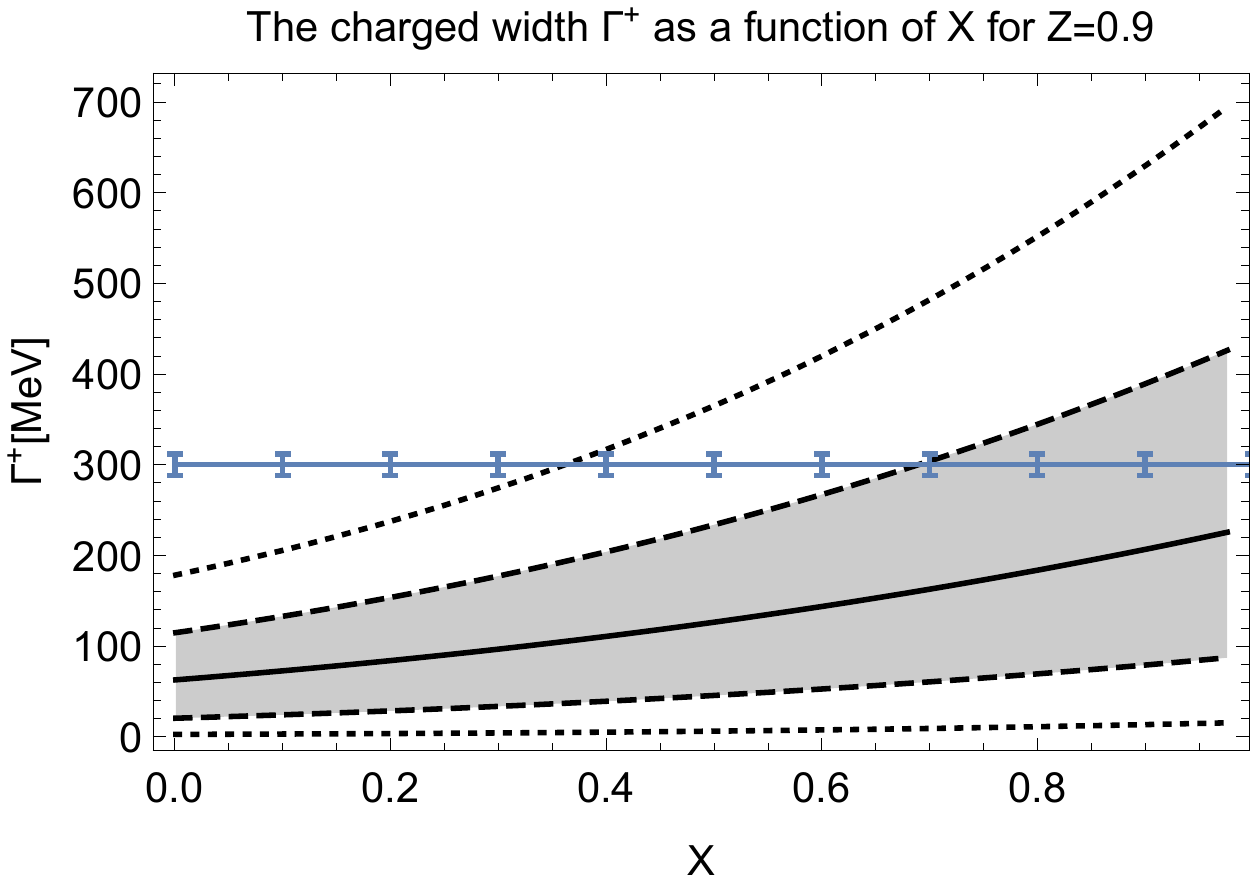}
\includegraphics[scale=0.6]{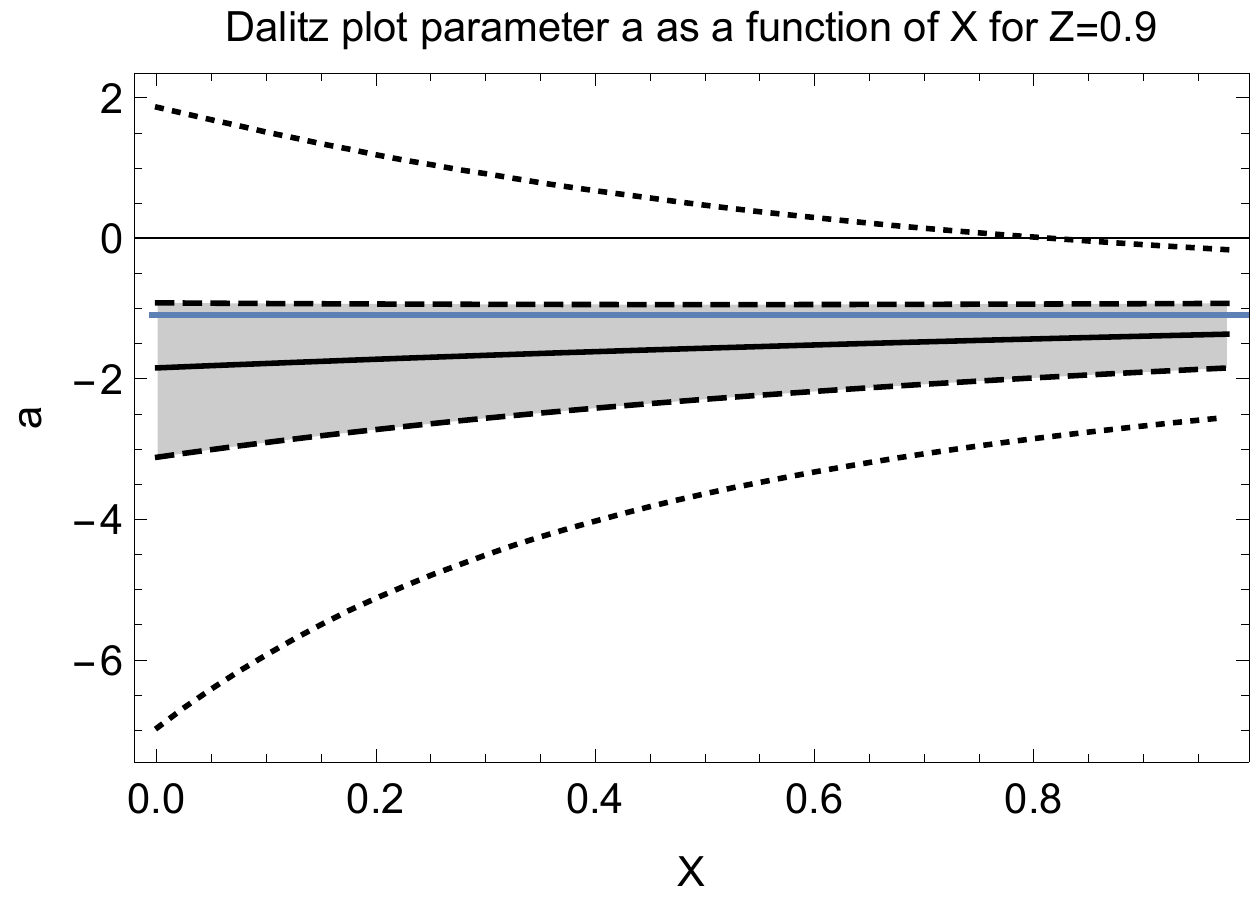}
\includegraphics[scale=0.6]{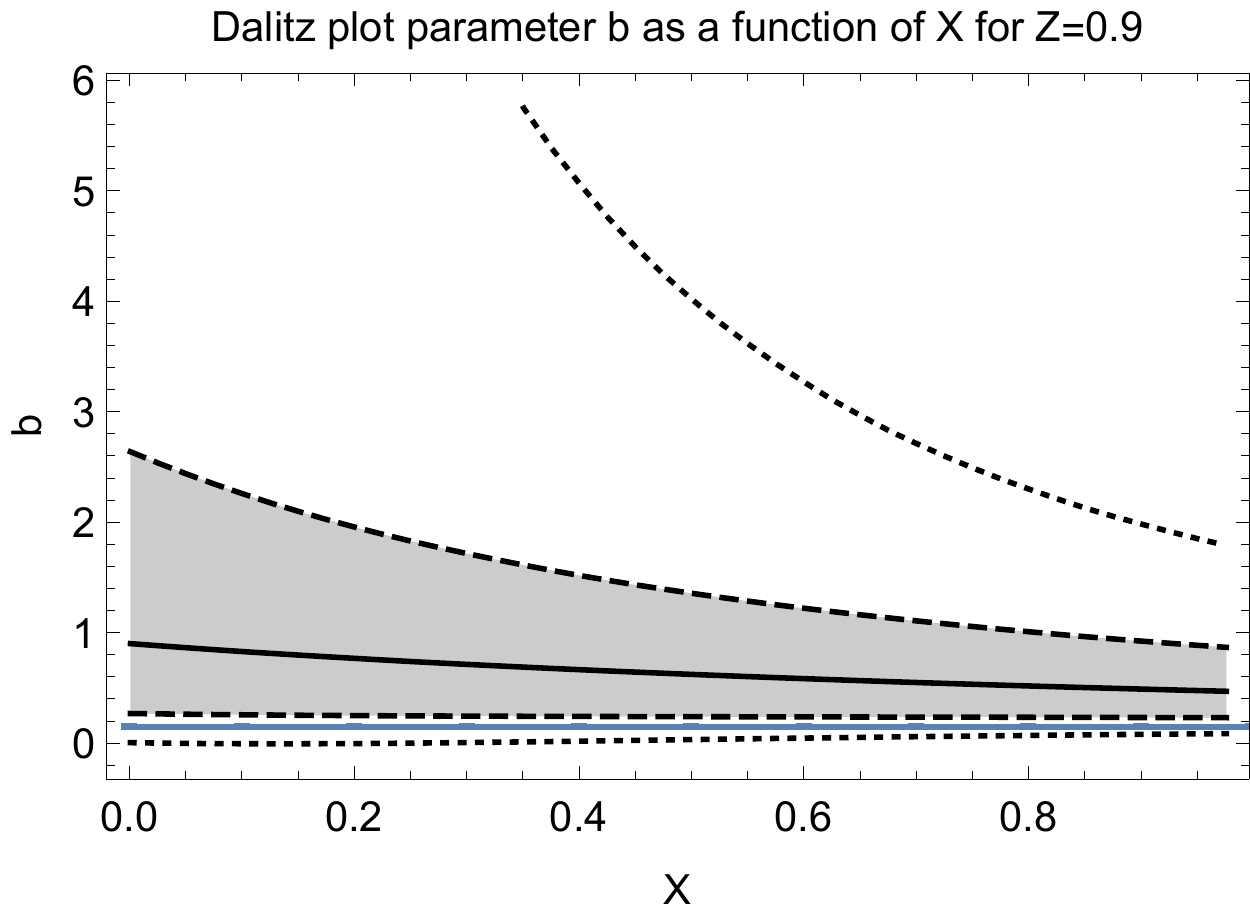}
\includegraphics[scale=0.6]{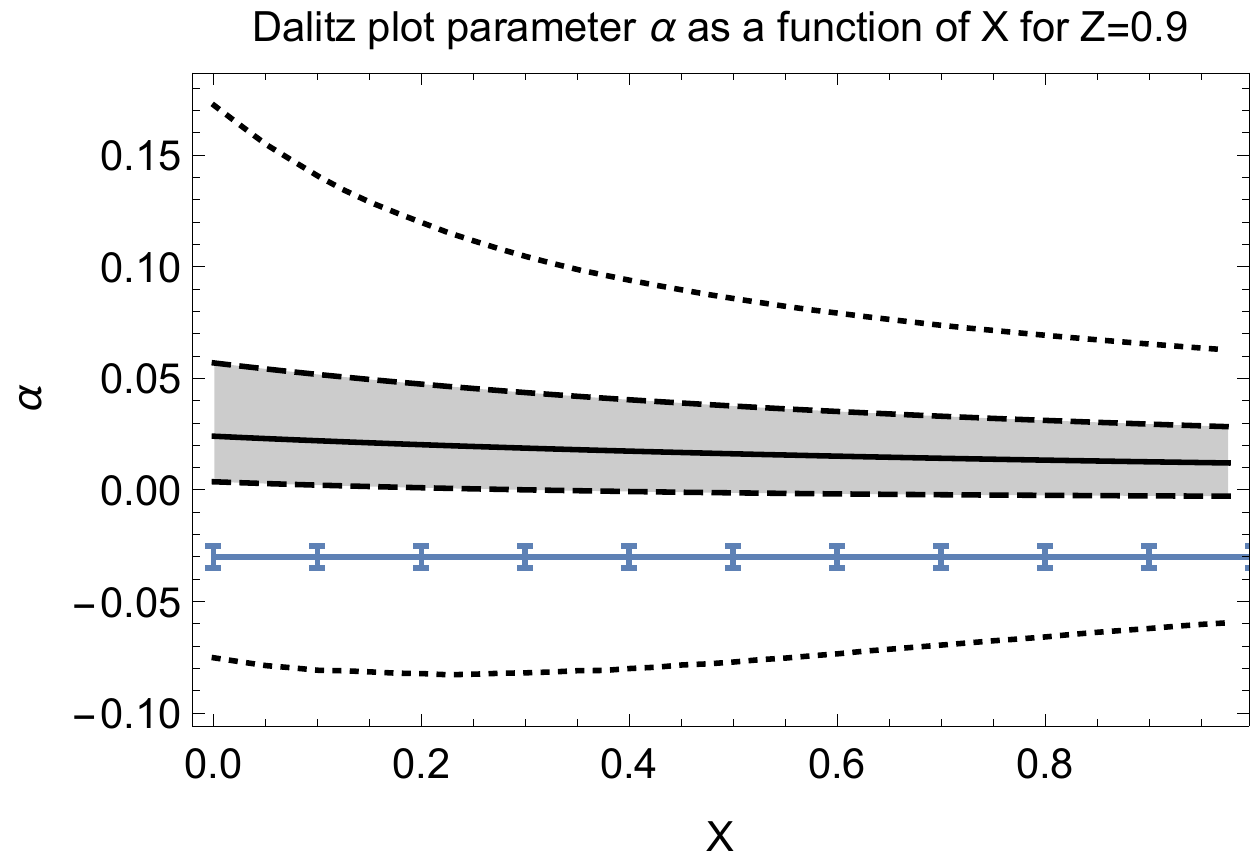}
\end{center}
\caption{Parameters $\Gamma^+$, $a$, $b$ and $\alpha$ as a function of $X$ for $Z=0.9$.\newline
The median (solid line), the one-sigma band (dashed,
shadowed) and the two-sigma band (dotted) are depicted along with the
experimental value \cite{PDG:2014kda,KLOE:2016qvh,KLOE:2010mj} (solid horizontal line with error bars).}
\label{f2}
\end{figure}

Full results can be found in \cite{Kolesar:2016jwe}, the main ones are reproduced here in figures \ref{f1} and \ref{f2}.

In the case of the decay widths, the experimental values can be reconstructed for a reasonable range of the free parameters and thus no tension is observed, in spite of what some of the traditional calculations suggest \cite{Osborn:1970nn,Gasser:1984pr}. As can be seen in figures \ref{f1} and \ref{f2}, we have found a strong dependence of the widths on $X$ and $Z$ and an appearance of both compatibility ($<1\sigma$ C.L.) and incompatibility ($>2\sigma$ C.L.) regions. Such a behavior is not necessarily in contradiction with the global convergence assumption and, moreover, it might be promising for constraining the parameter space and an investigation of possible scenarios of the chiral symmetry breaking \cite{Kolesar:2014zra}.

As for the Dalitz plot parameters, $a$ and $d$ can be described very well too, within $1\sigma$ C.L. As an example, results for $a$ are depicted in the figures.

However, when $b$ and $\alpha$ are concerned, we find a mild tension for the whole range of the free parameters, at less than 2$\sigma$ C.L. This marginal compatibility is not entirely unexpected. In the case of derivative parameters, obtained by expanding the amplitude in a specific kinematic point, in our case the center of the Dalitz plot, and depending on NLO quantities, the global convergence assumption is questionable, as discussed in \cite{Kolesar:2016jwe}. Also, the distribution of the theoretical uncertainties is found to be significantly non-gaussian, so the consistency cannot be simply judged by the 1$\sigma$ error bars.

The  marginal compatibility in the case of the parameters  $b$ and $\alpha$ can be interpreted in two ways - either some of the higher order corrections are indeed unexpectedly large or there is a specific configuration of the remainders, which is, however, not completely improbable.
 This warrants a further investigation of the higher order remainders by including additional information. Work is under way in analyzing $\pi\pi$ rescattering effects and resonance contributions, some preliminary results can be found in \cite{Kolesar:2011wn}.
\vsp{-0.25cm}
\\ \\
{\bf Acknowledgment:}\\
This work was supported by the Czech Science Foundation (grant no. GACR 15-18080S).

\vsp{-0.25cm}



%% References

\nocite{*}
\bibliographystyle{elsarticle-num}
\bibliography{Bibliography}

\begin{thebibliography}{10}
\expandafter\ifx\csname url\endcsname\relax
  \def\url#1{\texttt{#1}}\fi
\expandafter\ifx\csname urlprefix\endcsname\relax\def\urlprefix{URL }\fi
\expandafter\ifx\csname href\endcsname\relax
  \def\href#1#2{#2} \def\path#1{#1}\fi

\bibitem{Bose:1966}
S.~Bose, A.~Zimerman, Il Nuovo Cimento A 43 (1966) 1165--1167.

\bibitem{Bardeen:1967}
W.~A. Bardeen, L.~S. Brown, B.~W. Lee, H.~T. Nieh, Phys. Rev. Lett. 18 (1967)
  1170--1174.

\bibitem{Sutherland:1966zz}
D.~Sutherland, Phys.Lett. 23 (1966) 384.

\bibitem{Bell:1968mi}
J.~Bell, D.~Sutherland, Nucl.Phys. B4 (1968) 315--325.

\bibitem{Osborn:1970nn}
H.~Osborn, D.~Wallace, Nucl.Phys. B20 (1970) 23--44.

\bibitem{Weinberg:1978kz}
S.~Weinberg, Physica A96 (1979) 327.

\bibitem{Gasser:1983yg}
J.~Gasser, H.~Leutwyler, Annals Phys. 158 (1984) 142.

\bibitem{Gasser:1984gg}
J.~Gasser, H.~Leutwyler, Nucl.Phys. B250 (1985) 465.

\bibitem{Gasser:1984pr}
J.~Gasser, H.~Leutwyler, Nucl.Phys. B250 (1985) 539.

\bibitem{PDG:2014kda}
K.~A. Olive, et~al., Chin. Phys. C38 (2014) 090001.

\bibitem{Urech:1994hd}
R.~Urech, Nucl.Phys. B433 (1995) 234--254.
\newblock \href {http://arxiv.org/abs/hep-ph/9405341}
  {\path{arXiv:hep-ph/9405341}}.

\bibitem{Baur:1995gc}
R.~Baur, J.~Kambor, D.~Wyler, Nucl.Phys. B460 (1996) 127--142.
\newblock \href {http://arxiv.org/abs/hep-ph/9510396}
  {\path{arXiv:hep-ph/9510396}}.

\bibitem{Ditsche:2008cq}
C.~Ditsche, B.~Kubis, U.-G. Meissner, Eur.Phys.J. C60 (2009) 83--105.
\newblock \href {http://arxiv.org/abs/0812.0344} {\path{arXiv:0812.0344}}.

\bibitem{Bijnens:2007pr}
J.~Bijnens, K.~Ghorbani, JHEP 0711 (2007) 030.
\newblock \href {http://arxiv.org/abs/0709.0230} {\path{arXiv:0709.0230}}.

\bibitem{Kambor:1995yc}
J.~Kambor, C.~Wiesendanger, D.~Wyler, Nucl.Phys. B465 (1996) 215--266.
\newblock \href {http://arxiv.org/abs/hep-ph/9509374}
  {\path{arXiv:hep-ph/9509374}}.

\bibitem{Anisovich:1996tx}
A.~Anisovich, H.~Leutwyler, Phys.Lett. B375 (1996) 335--342.
\newblock \href {http://arxiv.org/abs/hep-ph/9601237}
  {\path{arXiv:hep-ph/9601237}}.

\bibitem{Colangelo:2011zz}
G.~Colangelo, S.~Lanz, H.~Leutwyler, E.~Passemar, PoS EPS-HEP2011 (2011) 304.

\bibitem{Kampf:2011wr}
K.~Kampf, M.~Knecht, J.~Novotny, M.~Zdrahal, Phys.Rev. D84 (2011) 114015.
\newblock \href {http://arxiv.org/abs/1103.0982} {\path{arXiv:1103.0982}}.

\bibitem{Guo:2015zqa}
P.~Guo, I.~V. Danilkin, D.~Schott, C.~Fernández-Ramírez, V.~Mathieu, A.~P.
  Szczepaniak, Phys. Rev. D92~(5) (2015) 054016.
\newblock \href {http://arxiv.org/abs/1505.01715} {\path{arXiv:1505.01715}}.

\bibitem{Bissegger:2007yq}
M.~Bissegger, A.~Fuhrer, J.~Gasser, B.~Kubis, A.~Rusetsky, Phys.Lett. B659
  (2008) 576--584.
\newblock \href {http://arxiv.org/abs/0710.4456} {\path{arXiv:0710.4456}}.

\bibitem{Gullstrom:2008sy}
C.-O. Gullstrom, A.~Kupsc, A.~Rusetsky, Phys.Rev. C79 (2009) 028201.
\newblock \href {http://arxiv.org/abs/0812.2371} {\path{arXiv:0812.2371}}.

\bibitem{Schneider:2010hs}
S.~P. Schneider, B.~Kubis, C.~Ditsche, JHEP 1102 (2011) 028.
\newblock \href {http://arxiv.org/abs/1010.3946} {\path{arXiv:1010.3946}}.

\bibitem{CrBarrel:1998yi}
A.~Abele, et~al., Phys.Lett. B417 (1998) 193--196.

\bibitem{KLOE:2008ht}
F.~Ambrosino, et~al., JHEP 0805 (2008) 006.
\newblock \href {http://arxiv.org/abs/0801.2642} {\path{arXiv:0801.2642}}.

\bibitem{KLOE:2016qvh}
A.~Anastasi, et~al., JHEP 05 (2016) 019.
\newblock \href {http://arxiv.org/abs/1601.06985} {\path{arXiv:1601.06985}}.

\bibitem{BESIII:2015cmz}
M.~Ablikim, et~al., Phys. Rev. D92 (2015) 012014.
\newblock \href {http://arxiv.org/abs/1506.05360} {\path{arXiv:1506.05360}}.

\bibitem{Wasa:2014aks}
P.~Adlarson, et~al., Phys. Rev. C90~(4) (2014) 045207.
\newblock \href {http://arxiv.org/abs/1406.2505} {\path{arXiv:1406.2505}}.

\bibitem{CrBarrel:1998yj}
A.~Abele, et~al., Phys.Lett. B417 (1998) 197--201.

\bibitem{Achasov:2001xi}
M.~Achasov, K.~Beloborodov, A.~Berdyugin, A.~Bogdanchikov, A.~Bozhenok, et~al.,
  JETP Lett. 73 (2001) 451--452.

\bibitem{CrBall:2001fm}
W.~Tippens, et~al., Phys.Rev.Lett. 87 (2001) 192001.

\bibitem{Wasa:2007aa}
M.~Bashkanov, D.~Bogoslawsky, H.~Calen, F.~Capellaro, H.~Clement, et~al.,
  Phys.Rev. C76 (2007) 048201.
\newblock \href {http://arxiv.org/abs/0708.2014} {\path{arXiv:0708.2014}}.

\bibitem{Wasa:2008vn}
C.~Adolph, et~al., Phys.Lett. B677 (2009) 24--29.
\newblock \href {http://arxiv.org/abs/0811.2763} {\path{arXiv:0811.2763}}.

\bibitem{CrBall:2008ny}
M.~Unverzagt, et~al., Eur.Phys.J. A39 (2009) 169--177.
\newblock \href {http://arxiv.org/abs/0812.3324} {\path{arXiv:0812.3324}}.

\bibitem{CrBall:2008ff}
S.~Prakhov, et~al., Phys.Rev. C79 (2009) 035204.
\newblock \href {http://arxiv.org/abs/0812.1999} {\path{arXiv:0812.1999}}.

\bibitem{KLOE:2010mj}
F.~Ambrosino, et~al., Phys.Lett. B694 (2010) 16--21.
\newblock \href {http://arxiv.org/abs/1004.1319} {\path{arXiv:1004.1319}}.

\bibitem{Fuchs:1991cq}
N.~Fuchs, H.~Sazdjian, J.~Stern, Phys.Lett. B269 (1991) 183--188.

\bibitem{DescotesGenon:1999uh}
S.~Descotes-Genon, L.~Girlanda, J.~Stern, JHEP 0001 (2000) 041.
\newblock \href {http://arxiv.org/abs/hep-ph/9910537}
  {\path{arXiv:hep-ph/9910537}}.

\bibitem{DescotesGenon:2003cg}
S.~Descotes-Genon, N.~Fuchs, L.~Girlanda, J.~Stern, Eur.Phys.J. C34 (2004)
  201--227.
\newblock \href {http://arxiv.org/abs/hep-ph/0311120}
  {\path{arXiv:hep-ph/0311120}}.

\bibitem{DescotesGenon:2007ta}
S.~Descotes-Genon, Eur.Phys.J. C52 (2007) 141--158.
\newblock \href {http://arxiv.org/abs/hep-ph/0703154}
  {\path{arXiv:hep-ph/0703154}}.

\bibitem{Kolesar:2016jwe}
M.~Kolesar, J.~Novotny\href {http://arxiv.org/abs/1607.00338}
  {\path{arXiv:1607.00338}}.

\bibitem{Bijnens:2014lea}
J.~Bijnens, G.~Ecker, Ann. Rev. Nucl. Part. Sci. 64 (2014) 149--174.
\newblock \href {http://arxiv.org/abs/1405.6488} {\path{arXiv:1405.6488}}.

\bibitem{Bijnens:1994ie}
J.~Bijnens, G.~Colangelo, J.~Gasser, Nucl.Phys. B427 (1994) 427--454.
\newblock \href {http://arxiv.org/abs/hep-ph/9403390}
  {\path{arXiv:hep-ph/9403390}}.

\bibitem{Amoros:2001cp}
G.~Amoros, J.~Bijnens, P.~Talavera, Nucl.Phys. B602 (2001) 87--108.
\newblock \href {http://arxiv.org/abs/hep-ph/0101127}
  {\path{arXiv:hep-ph/0101127}}.

\bibitem{Bijnens:2011tb}
J.~Bijnens, I.~Jemos, Nucl.Phys. B854 (2012) 631--665.
\newblock \href {http://arxiv.org/abs/1103.5945} {\path{arXiv:1103.5945}}.

\bibitem{Aoki:2013ldr}
S.~Aoki, et~al., Eur. Phys. J. C74 (2014) 2890.
\newblock \href {http://arxiv.org/abs/1310.8555} {\path{arXiv:1310.8555}}.

\bibitem{Kolesar:2014zra}
M.~Kolesar, J.~Novotny, Nucl. Part. Phys. Proc. 258-259 (2015) 90--93.
\newblock \href {http://arxiv.org/abs/1409.3380} {\path{arXiv:1409.3380}}.

\bibitem{Kolesar:2011wn}
M.~Kolesar, Nucl. Phys. Proc. Suppl. 219-220 (2011) 292--295.
\newblock \href {http://arxiv.org/abs/1109.0851} {\path{arXiv:1109.0851}}.

\end{thebibliography}






% \end{thebibliography}

\end{document}